\documentclass[twocolumn,preprintnumbers,amsmath,amssymb,floatfix]{revtex4}

\usepackage{graphicx}
\usepackage{dcolumn}
\usepackage{bm}
\usepackage{ifpdf}

\newif\ifcomment
\newif\ifprint
\printtrue

\ifpdf
\ifprint
 \usepackage[linkbordercolor={0 0 .8},%
             citebordercolor={.8 0 0},%
             urlbordercolor={.8 0 .8},%
             raiselinks=true,%
             pdfborder={0 0 1 [3]}]{hyperref}
\else
 \usepackage[pdftex,backref,pagebackref,colorlinks,
             linkcolor=blue,citecolor=blue,anchorcolor=blue,
             pagecolor=blue,urlcolor=red,a4paper]{hyperref}
\fi
\pdfoutput=1        
\pdfcatalog{/PageMode(/UseOutlines)} 
\else 
\ifprint
 \usepackage[linkbordercolor={0 0 .8},%
             citebordercolor={.8 0 0},%
             urlbordercolor={.8 0 .8}]
             {hyperref}
\else
 \usepackage[dvips,backref,pagebackref,colorlinks,
             linkcolor=blue,citecolor=blue,anchorcolor=blue,
             pagecolor=blue,urlcolor=red,a4paper]{hyperref}
\fi
\hypersetup{
  pdfauthor   = {},
  pdftitle    = {},
  pdfcreator  = {},
  pdfproducer = {},
  pdfsubject  = {},
  pdfkeywords = {}
 }
\fi

\newcommand {\abs}[1]    {\ensuremath{\left| #1 \right|}}
\newcommand {\der}{{\rm{d}}}

\newcommand {\vtwoatzero}{\mathcal{\scriptstyle V}_2}
\newcommand {\sigmatot}{\sigma_{\text{dyn}}}
\newcommand {\sigmavtwoatzero}{\sigma_{_{\vtwoatzero}}}

\newcommand {\meanvtwo}[1]{\bar{\vtwoatzero}^{\text{#1}}}

\begin{document}

\title{Event-by-event fluctuations of azimuthal particle anisotropy \\
  in Au+Au collisions at \mbox{\boldmath$\sqrt{s_{_{\it NN}}}$}=~200~GeV}

\author{
%
%
B.Alver$^4$,
B.B.Back$^1$,
M.D.Baker$^2$,
M.Ballintijn$^4$,
D.S.Barton$^2$,
R.R.Betts$^6$,
A.A.Bickley$^7$,
R.Bindel$^7$,
W.Busza$^4$,
A.Carroll$^2$,
Z.Chai$^2$,
M.P.Decowski$^4$,
E.Garc\'{\i}a$^6$,
T.Gburek$^3$,
N.George$^2$,
K.Gulbrandsen$^4$,
C.Halliwell$^6$,
J.Hamblen$^8$,
M.Hauer$^2$,
C.Henderson$^4$,
D.J.Hofman$^6$,
R.S.Hollis$^6$,
R.Ho\l y\'{n}ski$^3$,
B.Holzman$^2$,
A.Iordanova$^6$,
E.Johnson$^8$,
J.L.Kane$^4$,
N.Khan$^8$,
P.Kulinich$^4$,
C.M.Kuo$^5$,
W.Li$^4$,
W.T.Lin$^5$,
C.Loizides$^4$,
S.Manly$^8$,
A.C.Mignerey$^7$,
R.Nouicer$^{2,6}$,
A.Olszewski$^3$,
R.Pak$^2$,
C.Reed$^4$,
C.Roland$^4$,
G.Roland$^4$,
J.Sagerer$^6$,
H.Seals$^2$,
I.Sedykh$^2$,
C.E.Smith$^6$,
M.A.Stankiewicz$^2$,
P.Steinberg$^2$,
G.S.F.Stephans$^4$,
A.Sukhanov$^2$,
M.B.Tonjes$^7$,
A.Trzupek$^3$,
C.Vale$^4$,
G.J.van~Nieuwenhuizen$^4$,
S.S.Vaurynovich$^4$,
R.Verdier$^4$,
G.I.Veres$^4$,
P.Walters$^8$,
E.Wenger$^4$,
F.L.H.Wolfs$^8$,
B.Wosiek$^3$,
K.Wo\'{z}niak$^3$,
B.Wys\l ouch$^4$\\
\vspace{3mm}
\small
%
%
 $^1$~Physics Division, Argonne National Laboratory, Argonne, IL 60439-4843,
 USA\\
 $^2$~Physics and C-A Departments, Brookhaven National Laboratory, Upton, NY
 11973-5000, USA\\
 $^3$~Institute of Nuclear Physics PAN, Krak\'{o}w, Poland\\
 $^4$~Laboratory for Nuclear Science, Massachusetts Institute of Technology,
 Cambridge, MA 02139-4307, USA\\
 $^5$~Department of Physics, National Central University, Chung-Li, Taiwan\\
 $^6$~Department of Physics, University of Illinois at Chicago, Chicago, IL
 60607-7059, USA\\
 $^7$~Department of Chemistry, University of Maryland, College Park, MD 20742,
 USA\\
 $^8$~Department of Physics and Astronomy, University of Rochester, Rochester,
 NY 14627, USA\\
}

\begin{abstract}\noindent
  
  This paper presents the first measurement of event-by-event
  fluctuations of the elliptic flow parameter $v_2$ in Au+Au
  collisions at $\sqrt{s_{_{\it NN}}} =$ 200~GeV as a function of
  collision centrality.  The relative non-statistical fluctuations of
  the $v_2$ parameter are found to be approximately 40\%. The results,
  including contributions from event-by-event elliptic flow
  fluctuations and from azimuthal correlations that are unrelated to
  the reaction plane (non-flow correlations), establish an upper limit on
  the magnitude of underlying elliptic flow fluctuations.  This limit
  is consistent with predictions based on spatial fluctuations of the
  participating nucleons in the initial nuclear overlap region.  These
  results provide important constraints on models of the initial state
  and hydrodynamic evolution of relativistic heavy ion collisions.

\vspace{3mm}
\end{abstract}

\maketitle

Results from the Relativistic Heavy Ion Collider (RHIC) at Brookhaven
National Laboratory suggest that a dense state of matter is formed in
ultrarelativistic nucleus-nucleus collisions~\cite{WhitePaperBrahms,
WhitePaper, WhitePaperStar, WhitePaperPhenix}. Studies of final state
charged particle momentum distributions show that the produced matter
undergoes a rapid collective expansion transverse to the direction of
the colliding nuclei.  In particular, for collisions at non-zero impact
parameter, the expansion shows a significant anisotropy in the
azimuthal angle, strongly correlated with the anisotropic shape of the
initial nuclear overlap region. The dominant component of this
anisotropic expansion is called ``elliptic flow'' and is commonly
quantified by the second coefficient, $v_2$, of a Fourier
decomposition of the azimuthal distribution of observed particles
relative to the event-plane angle~\cite{Voloshin}.

Elliptic flow has been studied extensively in Au+Au collisions at RHIC
as a function of pseudorapidity, centrality, transverse momentum and
center-of-mass energy~\cite{PhobosFlowPRL1, PhobosFlowPRL2,
PhobosFlowPRC, WhitePaper, WhitePaperStar, WhitePaperPhenix}.  For Au+Au
collisions at RHIC energies, the observed dependence of the elliptic
flow signal on centrality and transverse momentum is found to be in
good agreement with calculations in hydrodynamic
models~\cite{HydroRef1, PhobosFlowPRC}.  This is considered evidence
for an early equilibration of the colliding system and a low viscosity
of the matter produced in the early stage of the collision
process~\cite{Teaney2003}.  In such calculations, for given conditions
of the produced matter, the elliptic flow magnitude is found
to be proportional to the eccentricity 
characterizing the transverse shape of the initial nuclear overlap
region~\cite{Ollitrault1992}.

Measurements of elliptic flow in the smaller Cu+Cu system have 
shown surprisingly large values of elliptic flow, in particular for
the most central collisions where the average eccentricity of the
nuclear overlap region was expected to be small~\cite{PhobosFlowPRL3}.
Experimental measurements of $v_2$ can be affected by event-by-event
fluctuations in the initial geometry~\cite{Miller} and it is possible
to reconcile the results for Cu+Cu and Au+Au collisions if these
fluctuations are properly accounted for~\cite{PhobosFlowPRL3}.
To this end, we have proposed a new definition of
eccentricity, which does not make reference to the direction of the
impact parameter vector, but rather characterizes the eccentricity
through the event-by-event distribution of nucleon-nucleon interaction
points obtained from a Glauber Monte-Carlo
calculation~\cite{PhobosFlowPRL3, PhobosGlauber}.  This method of
calculating the initial state anisotropy, which leads to finite
``participant eccentricity'' values even for the most central events
and has a large effect in the smaller Cu+Cu system, has been found to
be crucial for understanding the comparison of Cu+Cu and Au+Au
elliptic flow results~\cite{PhobosFlowPRL3}.

Using the probabilistic distribution of interaction points obtained
from a Glauber calculation, performed on an event-by-event basis,
leads to relative eccentricity fluctuations of
$\sigma_{\epsilon_{\text{part}}}/\langle \epsilon_{\text{part}}
\rangle\! \approx \!  40\%$ for Au+Au collisions at fixed number of
participating nucleons~($N_{\text{part}}$)~\cite{PhobosGlauber}.  Similar
calculations taking into account stochastic initial state interaction
points in a color glass condensate (CGC) model also yield large
relative eccentricity fluctuations of
$\sigma_{\epsilon_{\text{part}}}/\langle \epsilon_{\text{part}}
\rangle\! \approx \!  30\%$~\cite{cgc}.  If $v_2$ is proportional to
$\epsilon$, an event-by-event measurement of elliptic flow should
therefore exhibit sizable fluctuations in $v_2$, even at fixed
$N_{\text{part}}$.

An event-by-event measurement of the anisotropy in heavy ion collisions
is expected to yield fluctuations from three sources: statistical
fluctuations due to the finite number of particles observed, elliptic
flow fluctuations and other many-particle correlations. The statistical
fluctuations in the observed $v_2$ signal can be taken out with a
study of the measurement response to the input $v_2$ signal.  Particle
correlations other than flow~(non-flow correlations) such as HBT,
resonance decays and jets can resemble correlations due to elliptic
flow and have various effects on different flow measurements. In
particular, non-flow correlations can broaden the apparent $v_2$
distribution and enhance the observed $v_2$ fluctuations.  This letter
presents the first measurement of event-by-event dynamic fluctuations
in $v_2$, which include contributions from elliptic
flow fluctuations and non-flow correlations.

The data shown here were taken with the PHOBOS detector at RHIC during
the year 2004. The PHOBOS detector is composed primarily of silicon
pad detectors for tracking, vertex reconstruction, and multiplicity
measurements. Details of the setup and the layout of the silicon
sensors can be found elsewhere~\cite{PhobosDet}.
The collision trigger, event selection and centrality
determination are described in Ref.~\cite{pid63}.
The Monte Carlo~(MC) simulations of the detector performance are based
on the HIJING event generator~\cite{HIJING} and the GEANT
3.21~\cite{GEANT} simulation package, folding in the signal response
for scintillator counters and silicon sensors.

The PHOBOS multiplicity array, composed of single layer silicon pad
detectors, has a coverage of $\abs{\eta}<5.4$ over almost the full
azimuth.
We parametrize the pseudorapidity dependence, $v_2(\eta)$, with a
single parameter, $\vtwoatzero \equiv v_2(0)$, and a triangular or
trapezoidal shape, given by {$v^{\rm
    tri}_2(\eta)=\vtwoatzero\,(1-\frac{\abs{\eta}}{6})$}, or {$v^{\rm
    trap}_2(\eta) =
  \left\{^{\vtwoatzero\,\text{if}\,\abs{\eta}<2}_{\frac{3}{2}\, v^{\rm
        tri}_2(\eta)\,\text{if}\,\abs{\eta} \geq 2}\right.$},
respectively. 
Both of these parameterizations provide a reasonable description of
the measured pseudorapidity dependence of elliptic
flow~\cite{PhobosFlowPRC}.

The event-by-event measurement method has been developed to use all
the available information from the multiplicity array to measure the
elliptic flow at zero rapidity, $\vtwoatzero$, while allowing an
efficient correction for the non-uniformities in the acceptance.
Taking into account correlations only due to elliptic flow, the
probability of a particle with given pseudorapidity, $\eta$, to be
emitted in the azimuthal angle, $\phi$, in an event with elliptic flow
magnitude, $\vtwoatzero$, and event-plane angle $\phi_0$ is given by
\begin{equation}
  p(\phi | \vtwoatzero, \phi_0; \eta) = \frac{1}{2\pi}\left\{1+2
  v_2(\eta)\cos\left(2\left[\phi-\phi_{0}\right]\right)\right\}.
  \label{eq:defineprob}
\end{equation}
The direction of the event-plane angle, $\phi_0$, is expected to align
with the reaction plane angle if the initial geometry of heavy ion
collisions is defined by two smooth Wood-Saxon distributions or with
the participant eccentricity axis if initial geometry fluctuations are
indeed present.  In this measurement, $\phi_0$ is determined from the
distribution of final state particles without relying on any model
about the initial geometry of the collision.

The angular coordinates ($\eta,\phi$) of charged particles are
measured using the location of the energy deposited in the silicon
multiplicity detectors. After merging of signals in neighboring pads in cases
where a particle travels through more than a single pad, the deposited
energy is corrected for the angle of incidence, assuming that the
charged particle originated from the primary vertex.  Noise and
background hits are rejected by placing a lower threshold on this
angle-corrected deposited energy. Depending on $\eta$,
merged hits with less than 50-60\% of the energy loss expected for a
minimum ionizing particle are rejected~\cite{hitref}.  Since the
multiplicity array consists of single-layer silicon detectors, there
is no $p_{T}$, charge or mass information available for the
particles. All charged particles above a low-$p_{T}$ cutoff of about
7~MeV/c at $\eta$=3, and 35~MeV/c at $\eta$=0  
(the threshold below which a charged pion is
stopped by the beryllium beam pipe) are included on equal footing.  We
define the probability density function (PDF) for a hit position
$(\eta,\phi)$ for an event with $\vtwoatzero$ and event-plane angle
$\phi_0$ as
\begin{equation}
  P(\phi | \vtwoatzero, \phi_0; \eta) =
  \frac{1}{s(\vtwoatzero,\phi_{0};\eta)} p(\phi | \vtwoatzero, \phi_0;
  \eta),
\end{equation}
where the normalization parameter $s(\vtwoatzero,\phi_{0};\eta)$ is
calculated in small bins of $\eta$ such that the PDF folded with the
acceptance is normalized to the same value for different values of
$\vtwoatzero$ and $\phi_{0}$. The normalization parameter is given by
\begin{equation}
 s(v_2,\phi_{0},\eta) = \int_{\eta-\Delta\eta}^{\eta+\Delta\eta} A(\eta',\phi) p(\phi | v_2, \phi_0; \eta') \der \phi  \der \eta',
\end{equation}
where the acceptance function, $A(\eta,\phi)$ denotes the probability
of a particle moving in the $\eta,\phi$ direction to yield a
reconstructable hit.

For a single event, the likelihood function of $\vtwoatzero$ and
$\phi_0$ is defined as $L(\vtwoatzero,\phi_{0}) \equiv \prod_{i=1}^{n}
P(\phi_i | \vtwoatzero, \phi_0; \eta_i)$, where the product is over
all $n$ hits in the detector. The likelihood function describes the
probability of observing the hits in the event for the given values of
the parameters $\vtwoatzero$ and $\phi_0$. The parameters
$\vtwoatzero$ and $\phi_0$ are varied to maximize the likelihood
function and estimate the observed values, $\vtwoatzero^{\text{obs}}$
and $\phi_0^{\text{obs}}$, for each event.

The response of the event-by-event measurement is non-linear and
depends on the observed multiplicity $n$. Therefore, a detailed study
of the response function is required to extract the true $\vtwoatzero$
distribution from the measured $\vtwoatzero^{\text{obs}}$
distribution.  Let $f(\vtwoatzero)$ be the true $\vtwoatzero$
distribution for a set of events in a given centrality bin, and
$g(\vtwoatzero^{\text{obs}})$ the corresponding observed
distribution. The true and observed distributions are related by
\begin{equation}
  g(\vtwoatzero^{\text{obs}}) = \int
  K(\vtwoatzero^{\text{obs}},\vtwoatzero,n) \, f(\vtwoatzero) \, \der \vtwoatzero \, N(n) \, \der n,
  \label{eqkernel}
\end{equation}
where $N(n)$ is the multiplicity distribution of the given set of
events and $K(\vtwoatzero^{\text{obs}},\vtwoatzero,n)$ is the expected
distribution of $\vtwoatzero^{\text{obs}}$ for events with fixed input
flow $\vtwoatzero$, and constant observed multiplicity $n$.

The response function, $K(\vtwoatzero^{\text{obs}},\vtwoatzero,n)$ is
determined by performing the event-by-event analysis on modified
HIJING events with flow of fixed magnitude $\vtwoatzero$. The flow is
introduced by redistributing the generated particles in each event in
the $\phi$ direction according to the probability distribution given
by Eq.~\ref{eq:defineprob} and the assumed pseudorapidity dependence 
of $v_2$. For the two parameterizations of
$v_2(\eta)$, triangular and trapezoidal, used in the event-by-event
measurement, the corresponding response functions, $K^{\text{tri}}$
and $K^{\text{trap}}$, are calculated.  Fitting smooth functions
through the observed response functions decreases bin-to-bin
fluctuations and allows for interpolation in $\vtwoatzero$ and $n$.
The response of a perfect detector can be determined as a function of
event multiplicity~\cite{Ollitrault1992}.  In
practice, some empirical modifications to the ideal relation,
accounting for detector effects, significantly improve fits to the
response function, leading to
\begin{multline}
 K(\vtwoatzero^{\text{obs}},\vtwoatzero,n) = \frac{\vtwoatzero^{\text{obs}}}{\sigma^2} \, \\
\times 
     \exp  \left(-\frac{\left(\vtwoatzero^{\text{obs}}\right)^2+ \left( \vtwoatzero^{\text{mod}} \right)^2}{2\sigma^2}\right)
     I_{0} \left(\frac{\vtwoatzero^{\text{obs}} \vtwoatzero^{\text{mod}} }{\sigma^2}\right),
\end{multline}
with $\vtwoatzero^{\text{mod}}=(A\,n+B)\vtwoatzero$ and $\sigma =
C/\sqrt{n} + D$, and where $I_{0}$ is the modified Bessel function.
The four parameters~($A,B,C,D$) are obtained by fits to observed
$K(\vtwoatzero^{\text{obs}},\vtwoatzero,n)$ in the modified HIJING
samples.

Correcting for all known effects incorporated in our MC, we obtain the
true event-by-event $\vtwoatzero$ distribution, $f(\vtwoatzero)$,
which includes contributions from elliptic flow fluctuations and
non-flow correlations. We assume $f(\vtwoatzero)$ to be a
Gaussian in the range $\vtwoatzero>0$~\cite{footnote} with two
parameters, $\meanvtwo{}$ and $\sigmavtwoatzero$, denoting the mean and
standard deviation in the given range. For given values of the
parameters, it is possible to take the integral in Eq.~\ref{eqkernel}
numerically to obtain the expected $\vtwoatzero^{\text{obs}}$
distribution. Comparing the expected and observed distributions, the
values of $\meanvtwo{}$ and $\sigmavtwoatzero$ are found by a
maximum-likelihood fit. Midrapidity ($|\eta|\!<\!1$) results from the
two parameterizations of $v_2(\eta)$, triangular and trapezoidal, are
averaged to obtain the mean~$\langle v_2 \rangle = 0.5 (\frac{11}{12} \meanvtwo{tri} +
\meanvtwo{trap})$ and standard deviation~$\sigmatot = 0.5 (\frac{11}{12}
\sigmavtwoatzero^{\text{tri}} +
\sigmavtwoatzero^{\text{trap}})$ of the elliptic flow
parameter $v_2$, where the factor $\frac{11}{12}$
comes from integration over $\eta$.

The induced $v_2$ fluctuations arising from fluctuations in the number
of participating nucleons are calculated by parameterizing the
$\langle v_2\rangle$ versus $N_{\text{part}}$ results and folding them
with the $N_{\text{part}}$ distributions in each centrality bin.  The
relative contribution of these fluctuations to $\sigmatot$ is found
to be less than 8\%.  Results in this letter are presented after
subtraction of $N_{\text{part}}$ induced fluctuations.

\begin{figure}[t]
  \centering \includegraphics[width=0.47\textwidth]{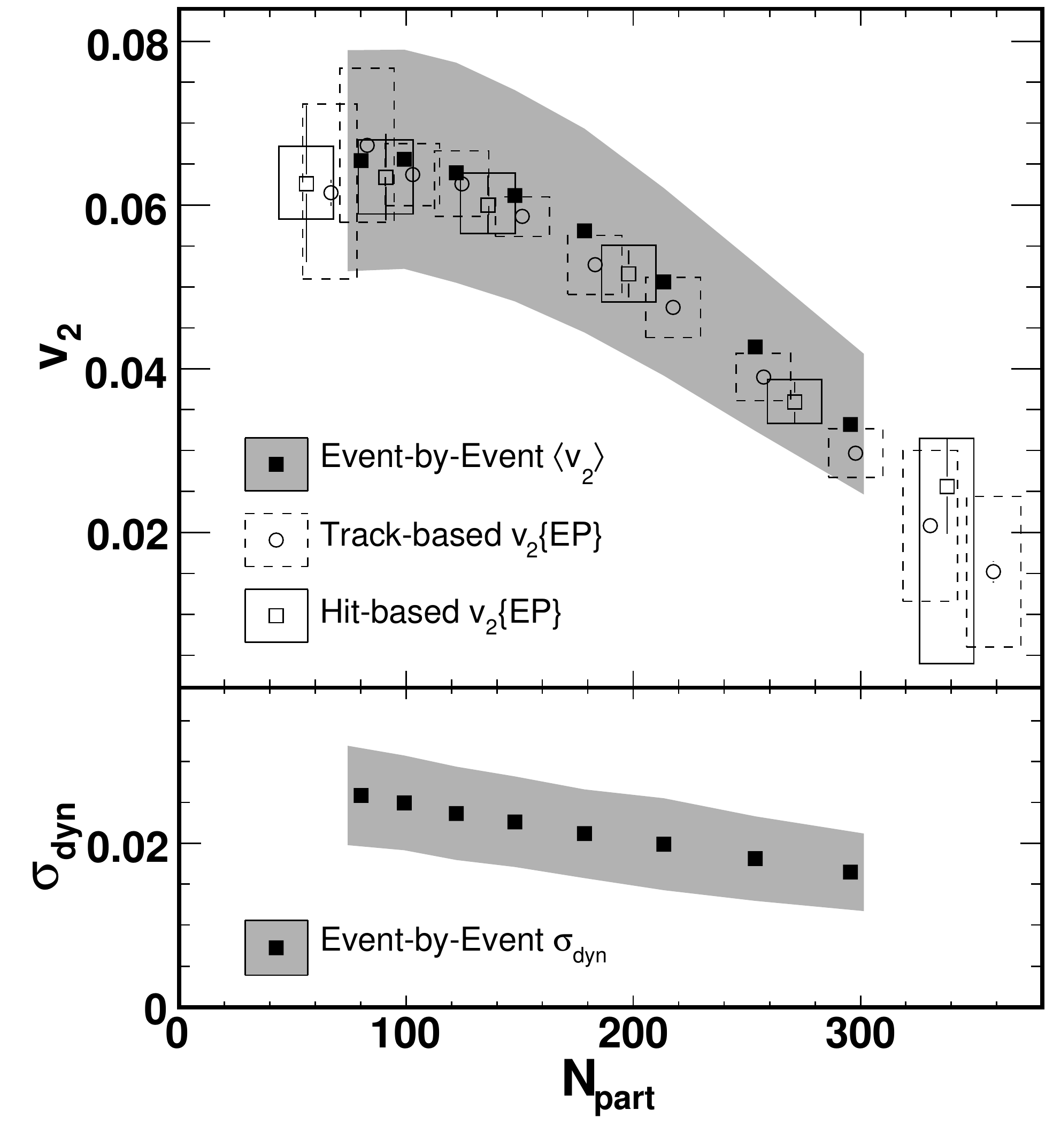}
  \caption { $\langle v_2 \rangle$ (top) and $\sigmatot$ (bottom)
    versus $N_{\text{part}}$ for Au+Au collisions at $\sqrt{s_{_{\it
          NN}}} =$ 200~GeV.  Previously published event-plane $v_2$
    results for the same collision system are shown for
    comparison~\cite{PhobosFlowPRL2}.  Boxes and gray bands show 90\%
    C.L.\ systematic errors and the error bars represent 1-$\sigma$
    statistical errors.  The results are for $0<\eta<1$ for the
    track-based method and $|\eta|<1$ for hit-based and event-by-event
    methods.}
  \label{fig:resultmeanrms}
\end{figure}

Systematic errors have been investigated in three main classes:
variations to the event-by-event analysis, response of the analysis procedure
to known input $\sigmatot$, and intrinsic differences between
HIJING events and data.  Various modifications to the event-by-event
analysis have been applied.  Corrections, previously used in the hit-based
event-plane analysis~\cite{PhobosFlowPRL1, PhobosFlowPRL2}, to account
for signal dilution due to detector occupancy and to create an
appropriately symmetric acceptance have been applied to both HIJING
and data events. The thresholds for background hit rejection have been 
varied.  These changes
lead to at most 4\% variations in the observed relative fluctuations
demonstrating a good understanding of the response function.  The
determination of the response function and the final fitting procedure
have been studied by performing the analysis on sets of modified
HIJING events with varying input $\sigmatot$.  Differences between
input and reconstructed $\sigmatot$ are identified as a
contribution to the systematic uncertainty.  The sensitivity of the
measurement is observed to be limited for very low $\langle
v_2\rangle$ values. Therefore the 0-6\% most central events, where the
reconstructed $\langle v_2\rangle$ is below 3\%, have been omitted.
Differences between HIJING and data in terms of $\der N/\der \eta$ and
$v_2(\eta)$ can, in principle, lead to a miscalculation of the
response function.  A sample of MC events has been generated, in which
the $\der N/ \der \eta$ distribution of HIJING events is widened by a simple
scaling to match the measurements in data within the
errors.  The difference between results obtained
with and without this modification, as well as the difference between
results with two different parameterizations of $v_2(\eta)$ are
identified as contributions to the systematic uncertainty.  Other
systematic studies include using a flat, rather than Gaussian, ansatz
for the true $\vtwoatzero$ distribution, $f(\vtwoatzero)$, and
performing the analysis in different collision vertex and event-plane
angle bins.  The uncertainty in the contribution of $N_{\text{part}}$
induced fluctuations has also been estimated via different
parameterizations of the $\langle v_2\rangle$ versus $N_{\text{part}}$
results.  Contributions from all error sources described above are
added in quadrature to derive the 90\% confidence level error.

\begin{figure}[t]
  \centering \includegraphics[width=0.47\textwidth]{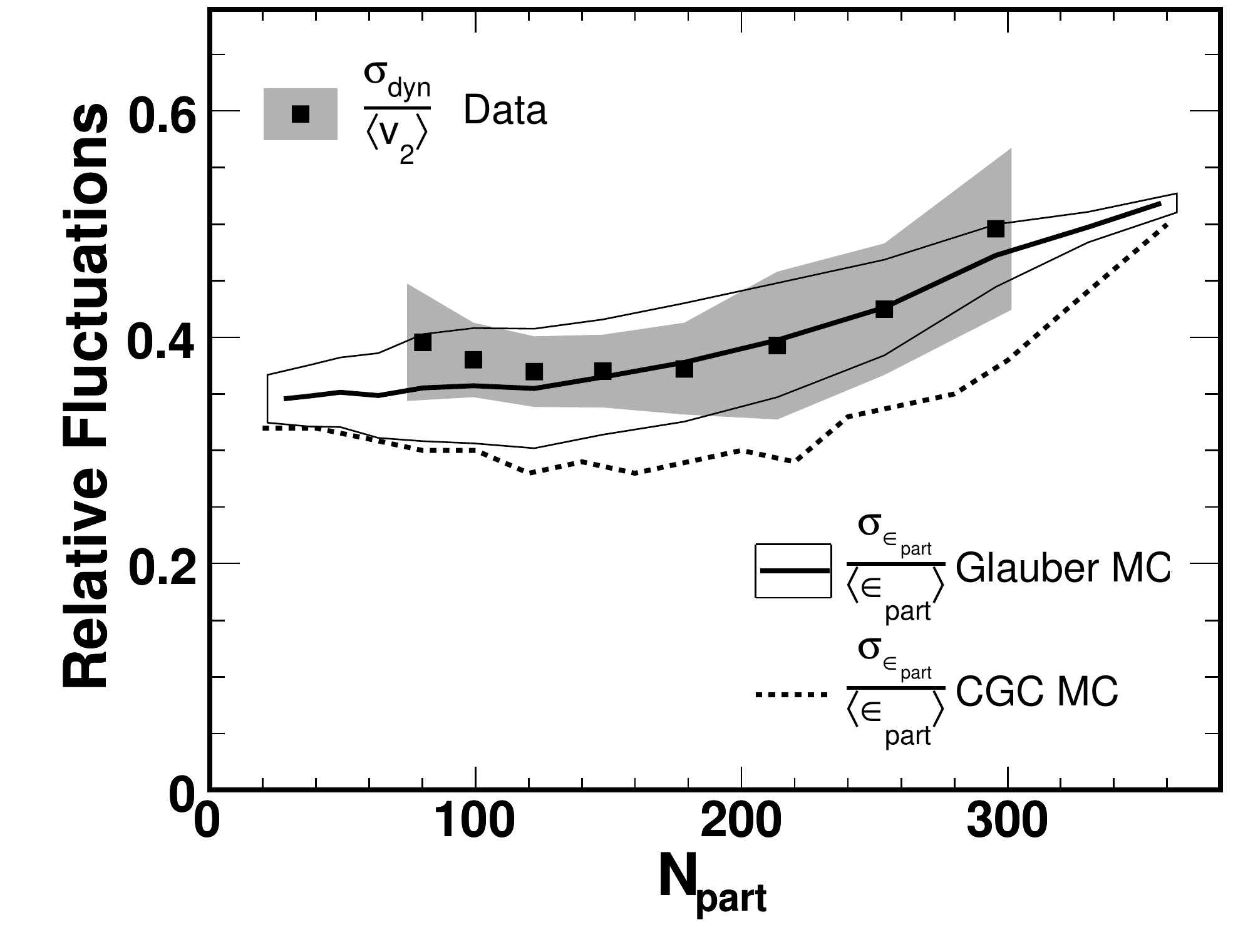}
  \caption{$\sigmatot/\langle v_2 \rangle$ versus $N_{\text{part}}$
    for Au+Au collisions at $\sqrt{s_{_{\it NN}}} =$ 200~GeV. The
    continuous and dashed thick black lines show
    $\sigma(\epsilon_{part})/\langle\epsilon_{part}\rangle$ calculated
    in Glauber MC~\cite{PhobosGlauber} and CGC~\cite{cgc} models,
    respectively.  The shaded grey band (for data) and thin black
    contour line (for Glauber MC) show 90\% C.L. systematics
    errors. See text for discussion of comparing the plotted data to
    the models.}
  \label{fig:resultwithecc}
\end{figure}

Fig.~\ref{fig:resultmeanrms} shows the mean, $\langle v_2 \rangle$,
and the standard deviation, $\sigmatot$, of the elliptic flow
parameter $v_2$ at midrapidity as a function of the number of
participating nucleons, in Au+Au collisions at $\sqrt{s_{_{\it NN}}}
=$ 200~GeV for 6--45\% most central events.  The results for $\langle
v_2 \rangle$ are in agreement with the previous PHOBOS $v_2$
measurements~\cite{PhobosFlowPRL2}, which were obtained with the
event-plane method for charged hadrons within \mbox{$|\eta| \! < \!
  1$}.  The uncertainties in $\der N/\der \eta$ and $v_2(\eta)$, as well as
differences between HIJING and the data in these quantities, introduce
a large uncertainty in the overall scale in the event-by-event
analysis due to the averaging procedure over the wide pseudorapidity
range.  The event-plane method used in the previous PHOBOS
measurements are known to be sensitive to the second moment,
$\sqrt{\langle v_2^2 \rangle}$, of elliptic flow~\cite{PhobosGlauber}.
The fluctuations presented in this letter would lead to a difference
of approximately 10\%
between the mean, $\langle v_2 \rangle$, and the RMS,
$\sqrt{\langle v_2^2 \rangle}$, of elliptic flow at a fixed value of
$N_{\text{part}}$.  However, a detailed comparison is not possible for
our $\langle v_2 \rangle$ measurements due to the scale errors, which
dominate the systematic uncertainty on $\langle v_2 \rangle$ and
$\sigmatot$.  Most of the scale errors cancel in the ratio,
$\sigmatot/\langle v_2 \rangle$, shown in Fig.~\ref{fig:resultwithecc}, 
revealing large relative fluctuations of approximately 40\%.

These results include contributions from both elliptic flow
fluctuations and non-flow correlations.  With no prior information on
the direction of the reaction plane, it is not possible to disentangle
these two contributions completely. However, several methods have been
proposed to estimate the contribution of non-flow correlations to the
observed dynamic $v_2$ fluctuations. One can assume that the
correlations in A+A collisions can be modeled by superimposing p+p
collisions~\cite{OlliFluc}. However, data from RHIC reveal many
differences between the overall correlation structure in Au+Au and p+p
(e.g.~\cite{Trainor,Wei,Starridge, Ed}). A more data-driven approach
assumes that non-flow correlations will be small for particle pairs
with large pseudorapidity separations (for example,
\mbox{$\Delta\eta>2$})~\cite{nonflow}. Under this latter assumption,
it is estimated that the relative fluctuations in the actual elliptic
flow account for a very large fraction (79-97\%) of the observed
relative dynamic fluctuations in the $v_2$ parameter~\cite{nonflow}.
No attempt was made to correct the data in
Fig.~\ref{fig:resultwithecc} for non-flow effects since the
validity of the large $\Delta\eta$ assumption cannot be unambiguously
tested with existing data.

The measured dynamic fluctuations in $v_2$ are directly comparable to
models that incorporate both elliptic flow and two particle
correlations. Furthermore, without making any assumptions about
non-flow, these data establish an upper limit on the magnitude
of underlying elliptic flow fluctuations. Also shown in
Fig.~\ref{fig:resultwithecc} are $\sigma_{\epsilon_{{\rm
      part}}}/\langle \epsilon_{{\rm part}} \rangle$ at fixed values
of $N_{part}$ obtained in MC Glauber~\cite{PhobosGlauber} and color glass
condensate(CGC)~\cite{cgc} calculations.  The 90\% confidence level
systematic errors for MC Glauber calculations (shown as a countour line in
Fig.~\ref{fig:resultwithecc}) are estimated by varying Glauber
parameters as discussed in Ref.~\cite{PhobosFlowPRL3}. 
Due to the uncertainties in non-flow effects discussed previously, it
is not possible to conclude which of these two models is more
consistent with the measured dynamic $v_2$ fluctuations.

In summary, we have presented the first measurement of event-by-event
$v_2$ fluctuations in Au+Au collisions at $\sqrt{s_{_{\it NN}}}
=$~200~GeV.  The relative non-statistical fluctuations of the $v_2$
parameter are found to be approximately 40\%. Independent estimates of
the non-flow correlation magnitude suggest that the major contribution
to these fluctuations are due to intrinsic elliptic flow fluctuations.
We show that the magnitude and centrality dependence of observed
dynamic fluctuations are consistent with predictions for fluctuations
of the initial shape of the collision region.  These results provide
qualitatively new information on the initial conditions of heavy ion
collisions and the subsequent collective expansion of the system.

%
%
%
%
This work was partially supported by U.S. DOE grants 
DE-AC02-98CH10886,
DE-FG02-93ER40802, 
DE-FG02-94ER40818,  
DE-FG02-94ER40865, 
DE-FG02-99ER41099, and
DE-AC02-06CH11357, by U.S. 
NSF grants 9603486, 
0072204,            
and 0245011,        
by Polish MNiSW grant N N202 282234 (2008-2010),
by NSC of Taiwan Contract NSC 89-2112-M-008-024, and
by Hungarian OTKA grant (F 049823).


\begin{thebibliography}{99}
\bibitem{footnote} The hit distribution on the detector due to
  negative and positive values of $\vtwoatzero$ are the same. Therefore a
  well-defined value for the fluctuations in $\vtwoatzero$ can only be
  obtained for $\vtwoatzero>0$. Any distribution with values $\vtwoatzero<0$ should be
  folded around zero for a proper comparison with the results in this
  paper.
\end{thebibliography}
\end{document}